\documentclass[12pt,epsfig]{article}

\usepackage{amssymb,amsmath,bm}
\usepackage{graphicx}

\usepackage[T2A]{fontenc}

\usepackage{epsfig,float}
\usepackage[hypertex,colorlinks=true,linkcolor=red,citecolor=blue]{hyperref}

\newcommand{\be}{\begin{eqnarray}}
\newcommand{\ee}{\end{eqnarray}}
\newcommand{\ba}{\begin{array}}
\newcommand{\ea}{\end{array}}

\newcommand{\bea}{\begin{eqnarray}}
\newcommand{\eea}{\end{eqnarray}}

\newcommand{\bi}{\begin{itemize}}
\newcommand{\ei}{\end{itemize}}

\newcommand{\nn}{\nonumber}

\begin{document}

\begin{titlepage}

\centerline{\large \bf
Quasi-Renormalizable Quantum Field Theories
}
\centerline{\large \bf
}

\vspace{5mm}

\centerline{
M.~V.~Polyakov$^{a,b}$,
K.~M.~Semenov-Tian-Shansky$^{b,c }$,}
\centerline{A.~O.~Smirnov$^d$,
and
A.~A. Vladimirov$^{e}$  }

\vspace{6mm}

\centerline{\it $^a$
Ruhr-Universit\"at Bochum, Fakult\"at f\"ur Physik und Astronomie, Institut f\"ur Theoretische
Physik II}
\centerline{\it  DE-44780 Bochum, Germany}

\vspace{4mm}

\centerline{\it $^b$
National Research Centre ``Kurchatov Institute'': Petersburg Nuclear Physics
Institute, }
\centerline{\it RU-188300 Gatchina, Russia}

\vspace{4mm}

\centerline{\it $^c$ Saint Petersburg National Research Academic University
of the Russian Academy of Sciences }
\centerline{\it RU-194021 Saint Petersburg, Russia}

\vspace{4mm}

\centerline{\it $^d$ Saint-Petersburg State University of Aerospace Instrumentation,}
\centerline{\it RU-190000, Saint Petersburg, Russia}

\vspace{4mm}

\centerline{\it $^e$ Universit\"at Regensburg, Institut f\"{u}r Theoretische Physik,  }
\centerline{\it DE-93040, Regensburg, Germany}

\vspace{4mm}

\centerline{\bf Abstract}
{ Leading
logarithms (LLs) in massless non-renormalizable effective field theories (EFTs)
can be computed with the help of non-linear recurrence relations. These recurrence relations
follow from the fundamental requirements of unitarity, analyticity and crossing symmetry of scattering
amplitudes and generalize the renormalization group technique for the case of non-renormalizable EFTs.
We review the existing exact solutions
of non-linear recurrence relations  relevant for  field theoretical applications.
We introduce the new class of quantum field theories (quasi-renormalizable field theories)
in which the resummation of LLs for $2 \to 2$ scattering amplitudes gives rise to a possibly
infinite number of the Landau poles.
 }

\noindent

\vspace{0.5cm}

\noindent

\noindent

\end{titlepage}

\section{Introduction}
\mbox

In Quantum Field Theories (QFTs) large logarithms of energy variables
usually occur in the calculation of loop corrections within the perturbation
theory approach. The presence of these large logarithmic corrections makes
it necessary to modify the original perturbation theory series in order to
ensure the consistency of the perturbative expansion. The standard tool
to address this issue is the Renormalization Group (RG) technique
(for a review see {\it e.g.}
Ref.~\cite{Vasiliev_green_Book}).
To master the so-called Leading Logarithms (LLs) (defined as the highest power
of a large logarithm at a given order of expansion in the coupling constant)
it suffices to take into account the result of a
one-loop calculation.
The RG-invariance makes it possible to partly take into account the perturbation theory
result to an infinitely large order of loop expansion by means of switching to the scale dependent running coupling constant. This improves the initial
perturbation theory series. By proper choice of energy scale one can tame the large
logarithmic corrections and thus broaden the applicability range of the
perturbation theory expansion.

An early attempt of systematic development of the RG technique for the case of
Effective Field Theories (EFTs) was performed by G.~Colangelo
\cite{Colangelo:1995np}
and also, in a somewhat different perspective, by D.~Kazakov
\cite{Kazakov:1987jp}.
The detailed formulation of the RG approach for the EFT case was given by
M.~Buchler and G.~Colangelo in
Ref.~\cite{Buchler:2003vw}.
M.~Bissenger and A.~Fuhrer in
Ref.~\cite{Bissegger:2006ix}
exploited these ideas and made extensive use of the analyticity requirements
to compute leading infrared%
\footnote{The term ``infrared'' refers to the low energy behavior: $E^2 \ll \mu^2$,
where $\mu$ stands for the characteristic theory scale.} logarithms of the  $\pi \pi$-scattering partial waves to
the three-loop accuracy in the massless
${\rm O}(4)/{\rm O(3)}$
theory. A major improvement was achieved in
Ref.~\cite{Kivel:2008mf}.
It was demonstrated that the RG invariance allows to compute the infrared LLs for the
$\pi \pi$-scattering amplitudes in the
${\rm O}(N)$-type%
\footnote{In the following we refer to $O(N+1)/O(N)$ and
${\rm SU}(N) \times {\rm SU}(N)/ {\rm SU}(N)$ sigma-models as for
${ \rm O}(N)$- and ${\rm SU}(N)$-type models respectively.} models
to an arbitrary high loop order. In
Ref.~\cite{Koschinski:2010mr} this result was reestablished
from the fundamental QFT requirements  of unitarity, analyticity and crossing
symmetry of scattering amplitudes. In Ref.~\cite{Kivel:2009az} a way to compute LLs to all orders of loop expansion
for scalar and vector form factors was pointed out in ${\rm O}(N)$-type models.
Further generalization of these ideas and detailed
studies of massless
${\rm SU}(N)$-type
models
were performed in the PhD thesis of A.~Vladimirov
\cite{VladimirovThesis}.
In Ref.~\cite{Polyakov:2010pt}
the non-linear recursion equations
for 
leading 
logarithms were generalized for the $4D$-sigma-model
with fields on an arbitrary Riemann manifold.
Recently, some new results on the LL-coefficients for the scalar and vector
form factors and the two point functions were presented in
Ref.~\cite{Ananthanarayan:2018kly}.

It worths mentioning that, contrary to the case of renormalizable theories,
in non-renormalizable EFTs the LLs do not fix the leading asymptotic behavior
of the Green functions since the logarithmic terms turn to be of the same order as
the non-logarithmic (polynomial) terms in dimensional variables. However, the special
interest in summing up these contributions can be argued from the impact parameter
space representation. In particular, LLs turn to be responsible for the large
$b_\bot$
asymptotic behavior of the
$3$D
(impact parameter dependent) parton distributions
$q(x, b_\bot)$
in the pion. Systematic accounting of these corrections results in the so-called
``chiral inflation'' of the pion radius
\cite{Perevalova:2011qi}.
This provides the explanation within the partonic picture  of the  logarithmic
divergency of the pion radius which is the familiar chiral perturbation theory
result.

Another motivation comes from the extensive studies of logarithmic
corrections in massive
${\rm O}(N)$- and ${\rm SU}(N)$-type
models \cite{Bijnens:2009zi,Bijnens:2010xg}
and in models involving fermionic degrees of freedom
\cite{Bijnens:2014ila}
performed by the group headed by J.~Bijnens.
The calculations in massless case provide useful consistency checks
for these computations.

In this paper, however, we would like to put forward a less rigorous
(though potentially more strong) motivation for study and systematic resummation
of LLs in massless EFTs.
Indeed, computing physical quantities to an arbitrary high order of loop expansion
can be seen as one of the major challenges in perturbative QFT.
This calculation can teach us important lessons on the general structure
of perturbation theory series and mirror some properties of the unknown
complete non-perturbative solution of the theory in question.
In case of usual renormalizable QFTs RG-logarithms
since long have been employed as a convenient test ground to implement this program.
We believe that the all-order resummation of LLs in non-renormalizable EFTs
can provide us valuable information of the perturbative loop expansion
in non-renormalizable EFTs  and will help to reveal non-trivial properties of
general non-perturbative amplitudes in these theories.

The paper is organized as follows. In
Sec.~\ref{Sec_2}
we review the main steps of the derivation of the recurrence relations for
the LL-coefficients of binary ($2 \to 2$) scattering amplitudes in massless
$\Phi^4$-type%
\footnote{The  lowest order term of the interaction Hamiltonian of such theory
involves $4$ field operators. } EFTs.
We present the general form of the  recurrence relations pertinent to the QFT
applications. We consider the possible singularities of the generating functions
of LL-coefficients and introduce the notion of quasi-renormalizable QFTs. These
QFTs can be seen as a generalization of the usual renormalizable QFTs.
In Sec.~\ref{Sec_3}
we present the known exact solutions of non-linear recurrence relations
relevant for the known examples of quasi-renormalizable QFTs.
Finally, our Conclusions are presented in
Sec.~\ref{Sec_Concl}.

\section{Non-linear Recurrence Relations for Leading Logarithms in Massless
$\Phi^4$-type EFTs}
\label{Sec_2}
\mbox

In this Section, following primary
Ref.~\cite{Koschinski:2010mr},
we review the key points of
derivation of the recurrence relations for
the coefficients of leading
logarithms of binary
scattering amplitudes of definite isospin in massless
$\Phi^4$-type EFTs
relying on the fundamental requirements of analyticity, unitarity
and crossing symmetry.

We consider a generic massless EFT with the following action:
\be
S=\int d^D x \left[ \frac{1}{2} \partial_\mu  \Phi^a  \partial^\mu  \Phi^a
- V(\Phi, \partial \Phi) \right].
\label{Action}
\ee
The Lagrangian
(\ref{Action})
is supposed to be invariant under some particular global group
$G$
which we refer to as ``isotopic group''.
$\Phi$ is the $N$-component vector in the isotopic space; its components  are
denoted by the index
$a$.
The interaction
$V(\Phi, \partial \Phi)$
is taken to
be of the
$\Phi^4$-type: the lowest order term is supposed
to involve $4$ field operators with
$2 \kappa$
derivatives. The parameter%
\footnote{So called chiral order of the interaction.}
$\kappa$
defines the dimension of the corresponding coupling constant
$1/F^2$: $[F]=\kappa+\frac{D}{2}-2$.

A prominent example of a theory belonging to the class
(\ref{Action}) is the
$O(N+1)/O(N)$
sigma-model:
\begin{eqnarray}
\label{ChPT}
&&
S=\int d^D x \frac{1}{2}\big(\partial_\mu\Sigma\partial^\mu\Sigma+\partial_\mu
\Phi^a
\partial^\mu
\Phi^a\big) \nn \\ && = \int d^D x \left(
\frac{1}{2}\partial_\mu\Phi^a\partial^\mu\Phi^a-\frac{1}{8F^2}(\Phi^a\Phi^a)\partial^2(\Phi^b\Phi^b)
+\mathcal{O}(\Phi^6) \right),
\end{eqnarray}
where
$\Sigma^2=F^2- \Phi^a\Phi^a$.
Here the isotopic group
$G$
turns to be
$O(N)$
and the chiral order of interaction is
$\kappa=1$.

Below, for simplicity, we provide explicit results
for the space-time dimension
$D=4$,
however the generalization for arbitrary even%
\footnote{The case of odd space dimensions is complicated by the
presence of non-logarithmic divergencies.}
space-time dimension $D>2$ is straightforward: see Apps.~A and B of
Ref.~\cite{Koschinski:2010mr}.
The special case of
$D=2$
is considered in details in
Ref.~\cite{Bissex}.

The main object of consideration is the  amplitude
$T_{abcd}(s,t,u)$
of the binary scattering process
$$
\Phi_a(p_1)+ \Phi_b(p_2) \to \Phi_c(p_3) + \Phi_d(p_4).
$$
Here
$s=(p_1+p_2)^2$, $t=(p_1-p_4)^2$ and $u=(p_1-p_3)^2$
are the usual Mandelstam variables satisfying
$s+t+u=0$
in the massless case. The amplitude is decomposed in  the irreducible
representations of the isotopic group
$G$
with the help of the corresponding projecting operators
${  P}_I^{abcd}$:
\be
T_{abcd}(s,t,u=-s-t)=\sum_I P^I_{abcd} T^{I}(s,t),
\ee
where
$I$
labels the appropriate irreducible representation of $G$.
The isotopic projectors
satisfy the completeness relation:
\be
\sum_I P^I_{abcd} = \delta_{ad}\delta_{bc}.
\label{completeness_rel}
\ee
The invariant amplitudes
$T^{I}(s,t,u)$
are further expanded into the Partial Waves (PWs) with respect to
the $s$-channel scattering angle in the center-of-mass system:
\be
T^I({s,t})=64 \pi \sum_{\ell=0}^\infty \frac{2 \ell +1}{2} P_\ell(\cos \theta_s) t_\ell^I(s).
\label{PW_expansion_D=4}
\ee
Here
$P_\ell(x)$
are the Legendre polynomials and the cosine of the
$s$-channel scattering angle is
$\cos \theta_s=1+\frac{2t}{s}$.
To the leading log accuracy, the $\ell$-th
PW-amplitude of the isospin
$I$
is given by
\be
t_{\ell}^I(s)= \frac{\pi}{2} \sum_{n=1}^\infty \frac{\hat{S}^n}{2\ell+1}
\omega_{n, \ell}^{I}
\ln^{n-1} \left( \frac{\mu^2}{|s|} \right)
+ {\cal O} ({\text{Next-to-Leading-Logs}}),
\label{PW_Llog_accuracy}
\ee
where
$\hat{S}=\frac{s^\kappa}{(4 \pi F)^2}$
is the dimensionless expansion parameter.
$\omega_{n, \ell}^{I}$
defined in (\ref{PW_Llog_accuracy})
are the 
LL-coefficients of the binary scattering PW-amplitudes, with the
index $n$ referring to the ``number of loops~$+1$'' and $\ell$
labeling the number of the PW.

The derivation of the recurrence relations for the LL-coefficients
$\omega_{n, \ell}^{I}$
in Ref.~\cite{Koschinski:2010mr}
relies on the fundamental requirements of unitarity, crossing symmetry and analyticity
of the PW amplitudes
$t_{\ell}^I(s)$
as the functions of the Mandelstam variable
$s$.
The right hand side cut discontinuity
($s>0$)
is, to the leading log accuracy, fixed by the
elastic ($2$-particle) unitarity relation:
\be
{\rm Disc } \, t_{\ell}^I(s)=|t_{\ell}^I(s)|^2+
{\cal O}({\rm Inelastic \ \ part} \sim \text{Next-to-Leading-Logs}).
\label{elastic_unitarity}
\ee
By means of the fixed-$t$ dispersion relation in the $s$-plane
and crossing symmetry
the left hand side cut discontinuity (for
$s<0$)
can be connected to right hand side discontinuity
by means of the generalization of the Roy equation
\cite{Bissegger:2006ix}:
\begin{eqnarray}
&&
{\rm Disc } \,t^I_{\ell}(s)=\sum_{\ell'=0}^\infty C_{su}^{IJ}\frac{2(2{\ell}'+1)}{s}\int_0^{-s}
ds' \,
P_{\ell}\Big(\frac{s+2s'}{-s}\Big)P_{{\ell}'}\Big(\frac{2s+s'}{-s'}\Big)
|t_{\ell'}^{J}(s')|^2 \nn
\\ && + {\cal O}({\rm Inelastic \ \ part} \sim \text{Next-to-Leading-Logs}),
\label{u_channel_elastic_unitarity}
\end{eqnarray}
where
$ C^{IJ}_{su}$
stand for the isospin crossing matrices connecting the invariant amplitudes
with interchanged momenta:
\be
T^{I}(s,t,u)= C^{IJ}_{su} T^{J}(u,t,s).
\ee
The crossing matrix
$C^{IJ}_{su}$
as well as the two complementary crossing matrices
$C_{st}^{IJ}$
and
$C_{tu}^{IJ}$
can be expressed through the projectors on the invariant  subspaces:
\be
C_{su}^{IJ}= \frac{1}{d_I} P^I_{abcd} P^{J}_{bdac}; \ \ \
C_{st}^{IJ}= \frac{1}{d_I} P^I_{abcd} P^{J}_{cbad}; \ \ \
C_{tu}^{IJ}= \frac{1}{d_I} P^I_{abcd} P^{J}_{bacd},
\label{Crossing_matrices}
\ee
where
$d_I=P^I_{abba}$
stand for the dimensions of the invariant subspaces.

Taking into the account the
$t \leftrightarrow u$
crossing symmetry together with the unitarity relations
(\ref{elastic_unitarity})
and
(\ref{u_channel_elastic_unitarity})
results in the following closed nonlinear recursive relation for the LL-coefficients:
\begin{eqnarray}
\omega^I_{n,{\ell}}=\frac{1}{n-1}\sum_J \sum_{k=1}^{n-1}
\sum_{{\ell}'=0}^{ \kappa n}\frac 12\ \Big(\delta^{{\ell}{\ell}'}\delta^{IJ}+
C_{st}^{IJ} \Omega_{\kappa n}^{{\ell}'{\ell}}
+C_{su}^{IJ}
(-1)^{{\ell}+
{\ell}'}\Omega_{\kappa n}^{{\ell}'{\ell}}\Big)\frac{\omega^{J}_{k,{\ell}'}
\omega^{J}_{n-k,{\ell}'}}{2{\ell}'+1}.
\label{RecRel_D=4_master}
\end{eqnarray}
Here the
$(\kappa n+1) \times (\kappa n+1)$
matrices
$\Omega_{\kappa n}^{{\ell'}{\ell}}$
defined through the relation
\be
\left( \frac{x-1}{2} \right)^{\kappa n} P_{\ell'} \left( \frac{x+3}{x-1} \right)= \sum_{\ell=0}^{\kappa n}
\Omega_{\kappa n}^{{\ell}'{\ell}}  P_{\ell}(x)
\label{Omega_mat_def}
\ee
perform the crossing transformation of the partial waves.
The initial conditions
$\omega^I_{n,{\ell}}$ with ${\ell}=0,\,1, \ldots\,, \kappa n$
for the recurrence relation
(\ref{RecRel_D=4_master})
are provided from the tree-level calculation of the PW amplitudes
$t^I_\ell(s)$
from the action
(\ref{Action}).

The recurrence relation
(\ref{RecRel_D=4_master})
allows to compute
the LL-coefficients to an arbitrary high loop order. It worths to emphasize that
the structure of these equations is  completely general for the $\Phi^4$-type theory:
the detailed form is defined
by the linear symmetry group
$G$ of the theory
(\ref{Action}) and the chiral order of the interaction
$\kappa$.
The details of the interaction also enter
through the initial conditions computed from the tree-level binary
scattering amplitude.

Numerical realization of the non-linear  recurrence relation
(\ref{RecRel_D=4_master})
was extensively employed for the calculation of the LL-coefficients for
various massless EFTs including the
${\rm O}(N+1)/{\rm O}(N)$-
and ${\rm SU}(N) \times {\rm SU}(N) /{\rm SU}(N)$-
sigma-models. The results were found to
be consistent with the finite loop order results known in the literature as well
as with the familiar results in the large-$N$ limit.

For the case of renormalizable QFT ($\kappa=0$ for $D=4$) the recurrence relation
(\ref{RecRel_D=4_master})
involves only the $\ell=0$ PW and
reduces to the much simpler form
\be
\omega_n^I= \frac{1}{n-1} \sum_{k=1}^{n-1} \sum_J {B}^{IJ}  \omega_k^J \omega_{n-k}^J,
\label{Rec_rel_renormalizable_th}
\ee
where the matrices
${B}^{IJ}$
are expressed in terms of the crossing matrices
(\ref{Crossing_matrices})
\be
{B}^{IJ}= \frac{1}{2} \left(\delta^{IJ}+C_{st}^{IJ}+ C_{su}^{IJ} \right)
\ee
and do  not depend on $n$. This makes the equation
(\ref{Rec_rel_renormalizable_th})
universal for any order of the loop expansion.
By introducing the generating function
\be
f^{I}(z)= \sum_{n=1}^\infty \omega^I_n z^{n-1} \ \ \
\text{with}
\ \ \ z \equiv \log \left( \frac{\mu^2}{|s|} \right),
\label{Gen_f_renormalizable}
\ee
where the parameter $\mu$ refers to the theory scale,
the non-linear recurrence relation can be put into the form
of the differential equation
\be
\frac{d \, f^I(z)}{dz}= \sum_J {B}^{IJ}  \left( f^{J}(z) \right)^2.
\label{RG_here}
\ee
This equation looks exactly like the RG equation for the running coupling constant
in the renormalizable QFT. In fact is no surprise, since in $\Phi^4$-type theories
the leading order evolution of the coupling is defined by the
$2 \to 2$ scattering amplitude.
Solving the equation (\ref{RG_here}) results in resummation of the
leading logarithmic corrections to all orders of loop expansion. The behavior of
the solution for the running coupling constant is determined by the presence of the Landau pole
\cite{Landau}.

It worths mentioning that in the case of renormalizable QFTs the  non-linear
recurrence relations similar to
(\ref{Rec_rel_renormalizable_th})
were also obtained in
Refs.~\cite{Malyshev:2003py,Malyshev:2004ax}
with the help of the RG-invariance formulation based on the properties of the
Lie algebra dual to the Hopf algebra of graphs \cite{Kreimer:1997dp}.

Moreover, recently the nonlinear recurrence relations  of the  similar form
were established to
perform the all-loop summation of the leading (and sub-leading)
divergencies for the binary scattering amplitudes in a class of maximally
supersymmetric gauge theories ($D=6,\,8,\,10$ supersymmetric Yang-Mills)
\cite{Borlakov:2016mwp,Kazakov:2016wrp,Borlakov:2017vra}.


\section{On the Simplified Form of the Recurrence Relations for LL-Coefficients}
\mbox

Finding the analytic solution of the recurrence relation
(\ref{RecRel_D=4_master})
would allow to sum up  leading logarithmic corrections to all orders of
loop expansion in the case of generic non-renormalizable massless $\Phi^4$-type EFT. 
Similarly to Eq.~(\ref{Gen_f_renormalizable}),
one can introduce the generating functions for
$\ell$-th partial wave amplitudes of
isospin
$I$:
\be
f^I_\ell(z) = \sum_{n=1}^\infty \omega^I_{n, \ell} z^{n-1}.
\label{Gen_f_I_ell}
\ee
However, for
$\kappa>0$
the equation
(\ref{RecRel_D=4_master})
looks almost unassailable for the analysis because of the strong mixing
between the different partial waves due to the presence of the
$\Omega^{\ell' \ell}_{\kappa n}$
matrices
(\ref{Omega_mat_def}).

Therefore, it is reasonable first to look for the solutions of the simplified
equations with reduced effect of mixing. In particular, such study can provide us
the necessary insight for the development of methods  to determine
the nature of the closest to the origin singularities of the generating functions
(\ref{Gen_f_I_ell}).
This would allow to  determine the large-$n$ asymptotic behavior of
$\omega^I_{n, \ell}$
and quantify the effect of resummation of LLs.

As pointed out in Ref.~\cite{Polyakov:2010pt}, certain useful simplification occurs in the
case of the
${\rm SU}(N)$-invariant theories in the large-$N$ limit.
The action of the   ${\rm SU}(N)$-invariant theory takes the form
\be
S=  \int d^4 x \, \frac{F^2}{4} \, {\rm tr} \, \left( \partial_\mu  U \partial^\mu U^\dag \right),
\label{SU(N)_action}
\ee
where $U=\exp [i \frac{\pi^a t^a}{F}]$ is the matrix of the Goldstone field. Here
$t^a$
are the generators of the
${\rm SU}(N)$ group
and the constant
$F$
is the lowest dimension coupling. The
chiral order of the interaction in
(\ref{SU(N)_action})
is
$\kappa=1$.

The first major simplification in the theory (\ref{SU(N)_action})
comes from the reduced mixing between the isospin
invariant subspaces.
The large-$N$ limit of the ${\rm SU}(N)$-symmetric theories turns
to be given by the interaction of just of two (out of $7$ possible)
subspaces
(symmetric and antisymmetric adjoint representations of
${\rm SU}(N)$: ${\rm Adj}_S$ and ${\rm Adj}_A$)
\be
\omega_{n, \ell}^{{\rm large}\, N}=
\left. \omega_{n, \ell}^{{\rm Adj}_S}\right|_{{\rm large}\, N}+\left.\omega_{n, \ell}^{{\rm Adj}_A}
\right|_{{\rm large}\, N}.
\ee
Moreover, for odd PWs
($\ell$ odd)
the LL-coefficients
$\omega_{n, \ell}^{{\rm large}\, N}$
are given by
\be
\left( \omega_{n, \ell}^{{\rm large}\, N}
\right)
= \frac{\rho_{n, \ell}}{16 F^2} \left( \frac{N}{2 F^2}\right)^{n-1},
\label{Expr_LL_SU(N)_l_odd}
\ee
where the coefficients $\rho_{n, \ell}$ satisfy a simpler
recurrence relation
\be
\rho_{n, \ell}= \frac{1}{2(n-1)} \sum_{k=1}^{n-1}
\sum_{\ell'=0}^n \frac{ \rho_{k, \ell'} \rho_{n-k, \ell'} }{2 \ell'+1}
\left(\delta^{\ell' \ell}+
{\Omega}_n^{\ell' \ell} \right)
\label{Rec_rel_rho_SU(N)}
\ee
with the initial conditions
$\rho_{1,0}=\rho_{1,1}=1$.
The expression for the LL-coefficients of even PWs ($\ell$ - even) is more bulky,
however finally they are expressed through the same coefficients
$\rho_{n, \ell}$
being the solution of the recurrence relations
(\ref{Rec_rel_rho_SU(N)}).

In Ref.~\cite{Julia} it was argued that for $\ell=0$ the
effect of the PW-mixing  in
(\ref{Rec_rel_rho_SU(N)})
turns to be negligible for large $n$.
This helps to simplify the form of the recurrence relation.
Another  promising possibility to get
rid of the complications due to the PW-mixing is to consider the case of two-dimensional
theory ($D=2$). The limited phase space in $D=2$ 
makes it possible only forward ($t=0$)
or backward ($u=0$) scattering. Therefore, the PW-mixing is reduced to a degenerate form involving
just forward and backward amplitudes. This issue is presented in details in Ref.~\cite{Bissex}.

This suggests the following simplified form of  the recurrence relation of real physical interest for
the LL-coefficients $f_n$:
\be
f_{n}=\frac{1}{n-1} \sum_{k=1}^{n-1} \, A(n,k) \, f_{k} f_{n-k},
\label{Phys_int}
\ee
where the function%
\footnote{The notation $A(n,k)$ is inspired by the Greek word
``${\cal A}\nu \alpha \delta \rho o \mu \dot{\eta}$'' for ``recursion''.}
 $A(n,k)$ encodes the properties of the LL-approximation of
the EFT in question.
Note that without loss of generality the initial condition for
(\ref{Phys_int})
can be taken as $f_{1}=1$.
To study the  recurrence relation
(\ref{Phys_int})  it is convenient to introduce the generating function
\be
f(z)= \sum_{n=1}^{\infty} f_n z^{n-1}; \ \ \ f(0)=1.
\label{Gen_f_master}
\ee

Mathematically, the problem of solving (\ref{Phys_int})
is related to the study of the non-linear integral equations of
the Hammerstein type
\cite{Hammerstein,Dolph}:
\be
f(z)- \int_a^b dy  K(z,y) f^2(y)=h(z),
\label{Hammer}
\ee
where $K(z,y)$ is a certain convolution kernel with specific
properties determined by the $A$-function
$A(n,k)$
and
$h(z)$
is a known function.
In some cases it turns out possible to reduce the integral equations
(\ref{Hammer})
to ordinary non-linear systems of differential equations for
the generating function
$f(z)$.

Our primary interest is
the asymptotic behavior of $f_n$ for large $n$.
It is determined
by the analytic properties of $f(z)$, particularly by the
position and nature of singularities closest to the origin.
Now we would like to introduce the class of
quasi-renormalizable QFTs that will be the main
subject of the present study.

{\bf Definition.}
{\it We call the quantum field theory  \textbf{quasi-renormalizable}
if the generating function
(\ref{Gen_f_master})
for the coefficients
of leading logs of binary scattering amplitude defined from the recurrence relation
of the type
(\ref{Phys_int})
is a meromorphic function of the variable
$z$.}

Note that the usual renormalizable QFTs match this definition
since in the latter case the generating function takes the form of a single
Landau pole. Therefore, quasi-renormalizable QFTs can be seen as a
certain generalization of usual renormalizable QFTs.

We would like to stress that the meromorphicity requirement in the whole
complex plane may turn
to be too much restrictive.
We may admit that
the generating function of the LL-coefficients
defined from the recurrence relation
(\ref{Phys_int})
in a sensible EFT may turn to be meromorphic only in a certain domain
in the complex plane ({\it e.g.} right half-plane).
Then, outside this domain, apart from poles it also can possess other types
of singularities ({\it e.g.} branching cuts).

In Ref.~\cite{Bissex} we present an example of quasi-renormalizable QFT
in $D=2$. This is the so-called O$(N)$-symmetric bissextile model with the
action
\be
&&
S= \int d^2x  \left( \frac{1}{2} \partial_\mu  {\Phi}^a \, \partial^\mu  {\Phi}^a -
g_1 (\partial_\mu  {\Phi}^a  \, \partial^\mu  {\Phi}^a)
(\partial_\nu  {\Phi}^b \, \partial^\nu  {\Phi}^b)-
g_2 (\partial_\mu  {\Phi}^a \, \partial^\nu  {\Phi}^a)
(\partial_\nu {\Phi}^b \, \partial^\mu  {\Phi}^b) \right), \nn \\ &&
\label{LagrangianBissextile}
\ee
where ${\Phi}$ is the $N$-component vector in the isotopic space and
$g_{1,2}$ stand for the coupling constants at the two possible vertices
involving $4$ derivatives. The term ``bissextile'' is used to emphasize
the fact that the four-point interaction in
(\ref{LagrangianBissextile})
contains the number of derivatives proportional to
$4$.
One can show that the non-linear recurrent system for the
LL-coefficients of binary scattering amplitudes in the theory
(\ref{LagrangianBissextile})
is equivalent to the following recurrence equation
\be
f_n=\frac{1}{n-1} \sum_{k=1}^{n-1}
\left(
A_0+A_1 (-1)^n+ A_2 (-1)^k
\right)
f_{n-k} f_k
\label{recrel_master1}
\ee
with the standard initial condition
$f_1=1$.
The values of the $A$-parameters are
\be
A_0= 1 + \frac{1}{(N+2)(N-1)}; \ \ \ A_1=-\frac{N+1}{(N+2)(N-1)};
\ \ \ A_2=-\frac{2}{(N+2)(N-1)}.
\ee
It is equivalent to the following functional differential equation for
the corresponding generating function (\ref{Gen_f_master}):
\be
f'(z)=A_0 f^2(z)+ A_1 f^2(-z)-A_2 f(z)f(-z); \ \ f(0)=1.
\label{Rec_rel_A0_A1_A2}
\ee
In Sec.~\ref{Sec_A0A1_equation} we consider eq.~(\ref{Rec_rel_A0_A1_A2})
for $A_2=0$ and present the existing solutions in terms
of elliptic functions (and their degeneracies).

The detailed analysis of the solutions of the recurrence relation
(\ref{recrel_master1})
for the bissextile model
(\ref{LagrangianBissextile})
is presented in Ref.~\cite{Bissex}. For several values of $N$,
meromorphic solutions were found for
(\ref{recrel_master1})
providing the living physical examples of quasi-renormalizable QFTs.

\section{Exact Solutions of Non-Linear Recurrence Relations}
\label{Sec_3}
\mbox

The analytic solution of non-linear recurrence relations
represent an extremely complicated mathematical problem due
to lack of reliable universal methods. 
In this Section we provide several examples of exact
solutions for particular recurrence relations of the type
(\ref{Phys_int})
and discuss the analytic properties of the corresponding generating
functions. In many cases the solutions of the recurrence systems that
share common features with recurrence systems for the QFT models turn to
possess meromorphic solutions.

\subsection{Case of Renormalizable Theory}
\mbox

First, for completeness, we consider
the recurrence relation
(\ref{Phys_int})
in the case
\be
A(n,k)=b_1= {\rm const}
\label{beta_factorial}
\ee
that
corresponds to  the usual renormalizable QFT. The constant $b_1$
is the one loop coefficient of the beta-function.
The recurrence relation
turns to be equivalent to the non-linear functional differential equation
\be
f'(z)= b_1 f^2(z); \ \ \ f(0)=1
\label{Difeq_renormalizable}
\ee
with the obvious solution
\be
f(z)=\frac{1}{1- b_1 z}
\ee
showing out the familiar Landau pole behavior.

\subsection{The Catalan Numbers}
\mbox

Another obvious example of the recurrence relation of the type
(\ref{Phys_int}) with
a non-trivial solution occurs for the choice
\be
A(n,k)=n-1.
\ee
The recurrence relation simply reads as
\be
f_n = \sum_{k=1}^{n-1} f_k f_{n-k},  \ \ \ f_1=1.
\label{rec_rel_Catalan}
\ee
The solution are the Catalan numbers $C_n$ (see {\it e.g.} \cite{Oesis_Cat})
\be
C_{n-1}= \frac{(2n-2)!}{n!(n-1)!} \equiv f_n.
\ee
These numbers admit plenty of combinatoric applications and shows up in various counting problems.
The generating function
(\ref{Gen_f_master})
for the Catalan numbers satisfies the following non-linear functional equation:
\be
f^2(z)= \frac{f(z)-1}{z}
\ee
and is expressed as
\be
f(z)=\frac{2}{1+\sqrt{1-4 z}}.
\ee
This function is obviously not meromorphic: it possesses a branching
point at
$z= \frac{1}{4}$.

\subsection{Solution in Terms of the Bessel Functions}
\mbox

A remarkable example of the recursive equation of type
(\ref{Phys_int})
was originally found by A.~Vladimirov. It corresponds to the
$A$-function chosen as
\be
A(n,k)=\frac{n-1}{n+\nu},
\label{Beta_Bessel}
\ee
where $\nu \ne -2,\,-3,\, \ldots$
is a parameter (and no dependence on $k$ is assumed).
The recurrence relation (\ref{Phys_int})  with the $A$-function
(\ref{Beta_Bessel})
turns to be equivalent to the following differential equation for the generating
function $f(z)$:
\be
f'(z)=f^2(z)- \frac{\nu+1}{z} (f(z)-f(0)); \ \ \ f(0)=1.
\ee
This equation can be linearized with the help of the substitution
$f(z)= \frac{u'(z)}{u(z)}$
and reduces to the  second order differential equation
\be
u''(z)+ \frac{\nu+1}{z} u'(z)-\frac{\nu+1}{z} u(z)=0.
\ee
The latter equation admits the general solution in terms of the modified Bessel functions
($I_\nu(x)=i^{-\nu} J_\nu(i x)$):
\be
&&
u(z)=z^{-\frac{\nu}{2}} (\nu+1)^{-\frac{\nu}{2}}
\left( {\cal C}_1 (-1)^{-\nu} I_{-\nu}(2 \sqrt{(\nu+1)z}) \Gamma(1-\nu) \right. \nn \\ &&
+\left. {\cal C}_2 I_{\nu}(2 \sqrt{(\nu+1)z}) \Gamma(1+\nu) \right).
\ee
Applying the boundary condition $\frac{u'(0)}{u(0)}=1$ to fix the values of the integration constants
${\cal C}_{1,2}$ after some algebra one finds
\be
f(z)= \sqrt{\frac{\nu+1}{z}} \frac{J_{\nu+1}(2 \sqrt{z (\nu+1)})}{J_{\nu }(2 \sqrt{z (\nu+1)})}.
\label{Lehas_Bessel_solution}
\ee
The solution
(\ref{Lehas_Bessel_solution})
possesses the remarkable analytic properties: for
$\nu>-1$
the function turn to be meromorphic in the right half-plane with poles along
$z>0$
axis.
The solution for
$\nu= \frac{1}{2}$
can be expressed in terms of the elementary trigonometric functions:
\be
f(z)= \sqrt{\frac{3}{2z}} \frac{J_{\frac{3}{2}}(\sqrt{6z })}{J_{\frac{1}{2}}(  \sqrt{6z })}=
\frac{1-\sqrt{6z} \cot \left(  \sqrt{6z}\right)}{2 z}.
\label{exact_sol_trig}
\ee
For
$\nu<-1$,
$f(z)$ (\ref{Lehas_Bessel_solution}) turns to be meromorphic in the left half-plane
with poles along
$z<0$.

It is interesting to mention that the ratio of the Bessel functions
similar to
(\ref{Lehas_Bessel_solution})
arose in the expression for the $2$-point function of the large-$N_c$ QCD
in the framework of the so-called meromorphization approach
developed by A.A.~Migdal
\cite{Migdal:1977nu,Migdal:1977ut,Migdal:2011pi,Migdal:2011gf}.
The meromorphization
procedure  consists in imposing proper analyticity
requirements for the set of the Green functions of a theory in order
to improve the convergence of the perturbative expansion.
It generalizes the Pad\'{e} approximation techniques in the limit of an infinite
order of approximation.
This allows to elaborate the set of highly non-trivial restrictions
for the physical mass spectrum and the anomalous dimensions of the
perturbation theory.
The additional motivation to revive the meromorphization approach of
A.A.~Migdal came from the observation that the same spectrum (roots of the Bessel
functions) can be derived from the AdS/CFT theory framework \cite{Erlich:2006hq}.

Also we would like to note that in Ref.~\cite{Borlakov:2016mwp}
the periodic solutions resembling much
Eq.~(\ref{exact_sol_trig})
occur for the generating function performing the summation of
all order leading divergencies coming from the ladder diagrams for
binary scattering amplitudes
in the $D=8$ and $D=10$ supersymmetric Yang-Mills theories
({\it cf.} Eqs.~(11) and (15) of Ref.~\cite{Borlakov:2016mwp}).

\subsection{Pseudofactorial and   Dixon's Elliptic  Functions}
\mbox

Now we discuss the highly non-trivial solution
\cite{Baher}
of the recurrence relation of the
type (\ref{Phys_int})
associated with the Dixonian and Weierstra{\ss} elliptic (meromorphic, doubly periodic) functions.
It occurs for
\be
A(n,k)=-(-1)^{n}.
\ee
In this case the recurrence relation (\ref{Phys_int})
\be
f_n = \frac{1}{n-1} \sum_{k=1}^{n-1} (-1)^{n-1} f_k f_{n-k},  \ \ \ f_1=1,
\label{Recrel_Pf}
\ee
defines the so-called pseudofactorial%
\footnote{The term ``pseudofactorial'' was introduced to emphasize the natural association
with the case of sign-non-altering $A$-function
(\ref{beta_factorial}).
In this instance the sequence of the usual factorials is generated through
(\ref{sequence}).}
 integer sequence \cite{Oesis_Pf}:
\be
\{ a_{n-1} \equiv (n-1)! f_n\}=  \{1,\, -1,\, -2,\, 2,\, 16,\, -40,\, -320,\, 1040,\,...\,\}.
\label{sequence}
\ee
The recurrence relation
(\ref{Recrel_Pf})
is equivalent to the following non-linear functional differential equation for the corresponding generating function
\be
f'(z)=-f^2(-z), \ \ \ f(0)=1.
\label{Diff_Eq_Bacher}
\ee
Note that the main difference with the equation
(\ref{Difeq_renormalizable})
corresponding to the usual renormalizable
theory case
is just the sign of the argument of  $f^2$ in the r.h.s. of
(\ref{Diff_Eq_Bacher}).

We follow Ref.~\cite{Baher} (see also Ref.~\cite{Flajolet})
to describe the solution of (\ref{Diff_Eq_Bacher}). It is convenient  to introduce the generating function of the
negative argument $g(z) \equiv f(-z)$.
The  non-linear functional differential equation (\ref{Diff_Eq_Bacher})
turns to be equivalent to the following system of differential
equations:
\be
\begin{cases}
f'(z)= -g^2(z), \\
g'(z) =f^2(z);
\label{System_Bacher}
\end{cases}
\ee
with the initial conditions $f(0)=1$, $g(0)=1$.
This system possesses a simple first integral:
\be
f^3(z)+g^3(z)=2.
\label{First_Int}
\ee
With its help the problem of finding the solution of (\ref{System_Bacher})
is reduced to the inversion of the following
integral:
\be
\int_{f(z)}^1 \frac{d y}{ (2-y^3)^{\frac{2}{3}}}=z.
\label{Bachers_f}
\ee

\begin{figure}[H]
 \begin{center}
 \epsfig{figure= 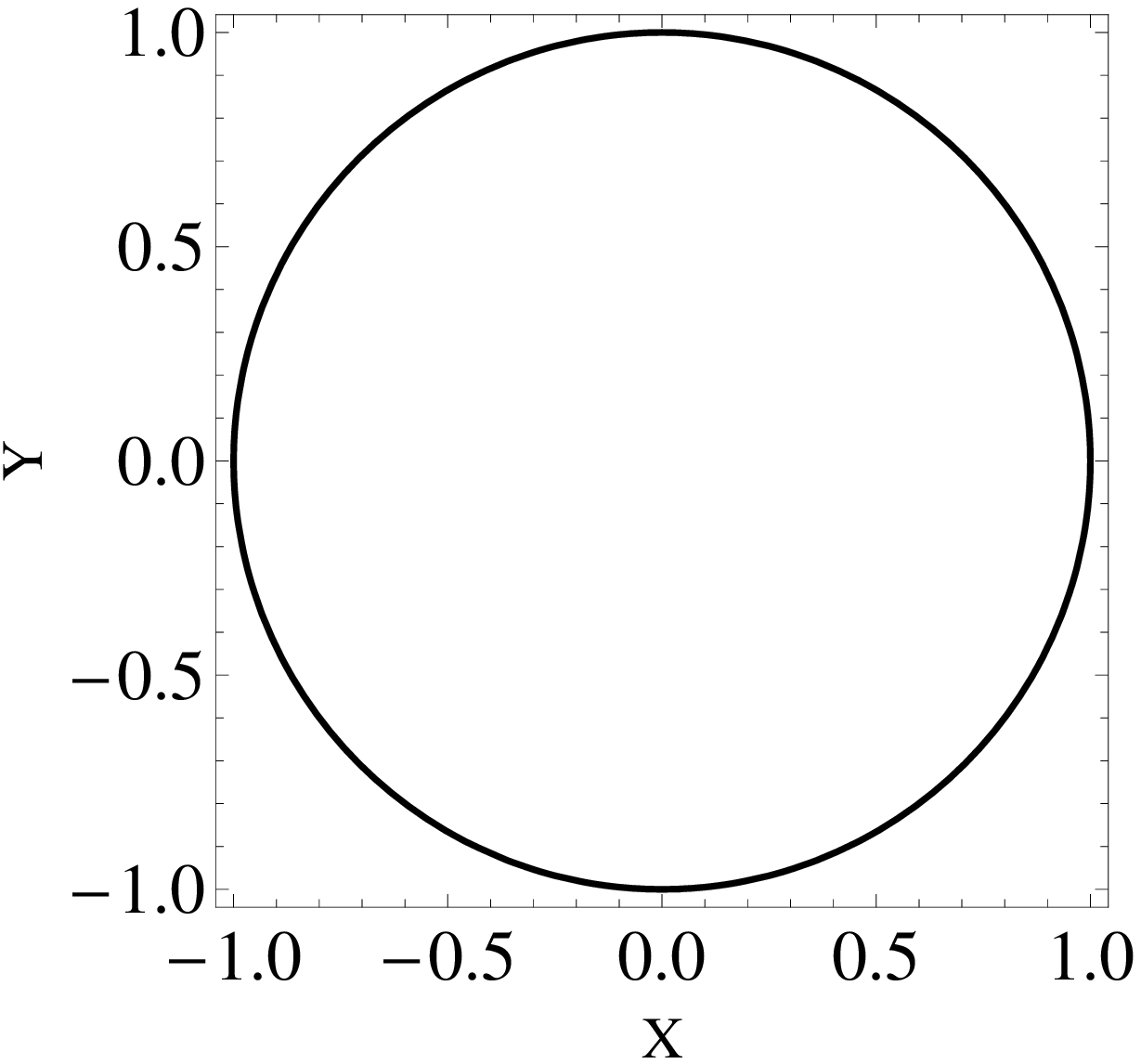 , height=7cm}
  \epsfig{figure= 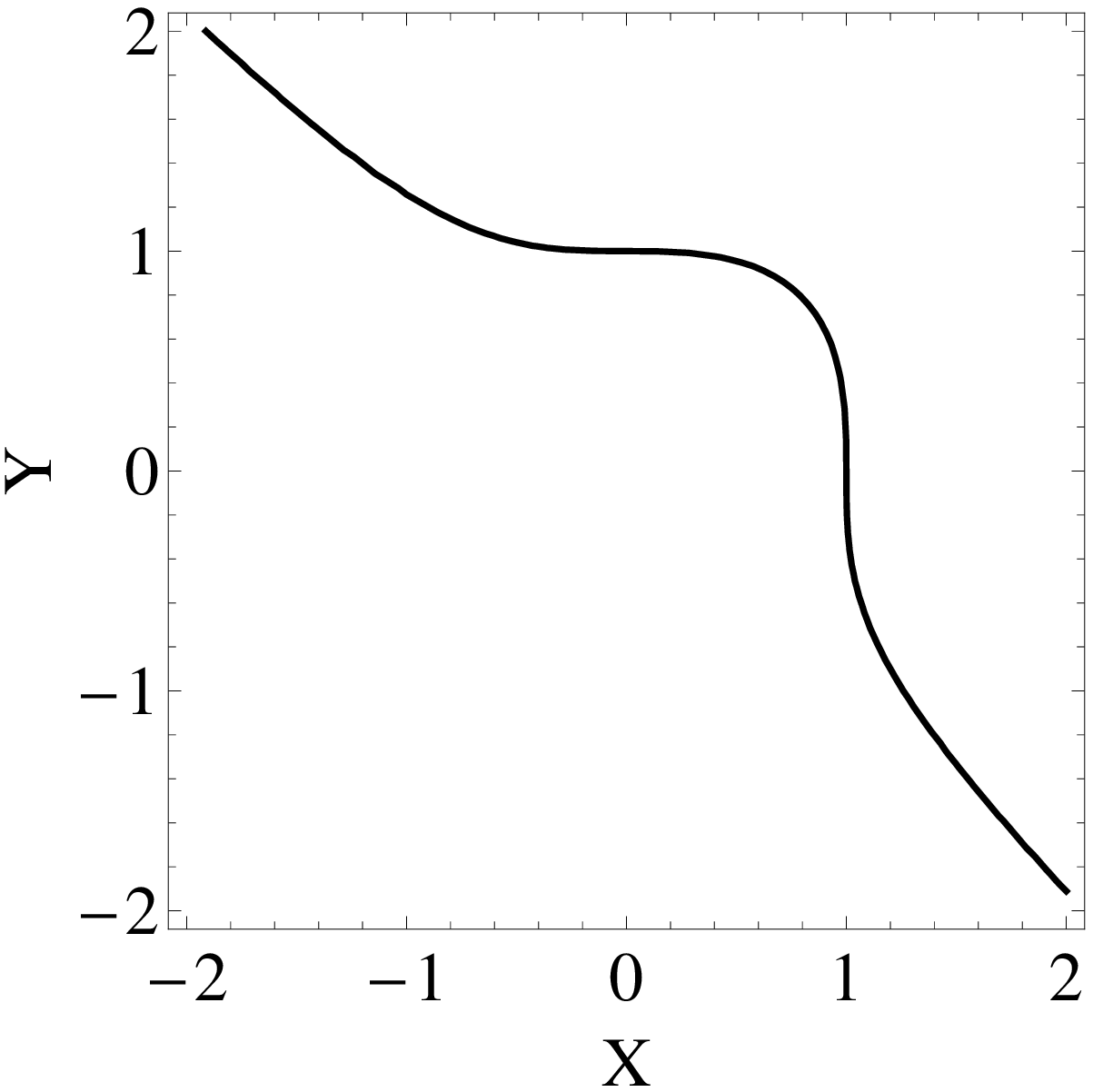 , height=7cm}
 \end{center}
  \caption{{\bf Left panel:} The circle  $X^2+Y^2=1$  defines usual sine and cosine functions. {\bf Right panel:} The Fermat cubic curve  $X^3+Y^3=1$  defines Dixon's elliptic functions (\ref{Dixon_functions}).}
\label{Fig_Fermat}
  \end{figure}

The function $f(z)$ can be expressed in terms of the so-called
Dixon's elliptic functions.
The pair of the functions
${\rm sm}$
and
${\rm cm}$ was introduced by A.~C.~Dixon in
\cite{Dixon}
and can be seen as a certain generalization of the conventional
sine and cosine functions respectively.
The functions
${\rm sm}$
and
${\rm cm}$
satisfy the
system of the differential equations
\be
\begin{cases}
{\rm sm}'(z)= {\rm cm}^2(z),  \\
{\rm cm}'(z) =-{\rm sm}^2(z);
\end{cases}
\ \ \ {\rm sm}(0)=0;  \ \ \ {\rm cm}(0)=1.
\label{Dixon_functions}
\ee
The following identity is valid:
\be
{\rm sm}^3(z) + {\rm cm}^3(z) =1.
\ee
Therefore, similarly to the usual sine and cosine defined by the circle
$X^2+Y^2=1$,
the pair
$({\rm sm}(z), {\rm cm}(z))$
parameterizes the Fermat cubic set by
the equation $X^3+Y^3=1$ (see Fig.~\ref{Fig_Fermat}).

Dixon's functions turn to be inverses of the 
integrals
\be
\int_0^{{\rm sm}z} \frac{dy}{(1-y^3)^{\frac{2}{3}}}=z; \ \ \
\int_{{\rm cm}z}^1  \frac{dy}{(1-y^3)^{\frac{2}{3}}}=z.
\ee
The real period of  Dixon's functions $\pi_3$
can be computed as
\be
\pi_3= 3 \int_{ 0}^{ 1} \frac{d y}{(1-y^3)^{\frac{2}{3}} }=
B \left (\frac{1}{3}, \, \frac{1}{3} \right)= \frac{\sqrt{3}}{2 \pi} \Gamma^3 \left( \frac{1}{3} \right),
\label{pi3}
\ee
where $B$ is the Euler beta function. Dixon's functions satisfy
\be
{\rm sm}(z+\pi_3)={\rm sm}(z); \ \ \
{\rm cm}(z+\pi_3)={\rm cm}(z); \ \ \ {\rm sm}\left(\frac{\pi_3}{3}-z \right) ={\rm cm}(z).
\ee
The complete lattice of periods is hexagonal and can described as
\be
a \pi_3 \oplus b \pi_3 e^{\frac{2 \pi i}{3}},
\ee
where $a$ and $b$ are arbitrary integers.

Finally, the solution of
(\ref{System_Bacher})
can be expressed as
\cite{Baher}:
\be
 f(z)=2^{\frac{1}{3}} {\rm sm} (\frac{\pi_3}{6}- 2^{\frac{1}{3}}z),
 \label{Pseud_Fact_f}
\ee
where
$\pi_3$
(\ref{pi3})
is the real period of the
${\rm sm}$
function.

To specify the analytic properties of
(\ref{Pseud_Fact_f})
it is convenient to express
$f(z)$
through the  Weierstra{\ss} elliptic $\wp$-function.
For this issue we divide $f(z)$ and $g(z) \equiv f(-z)$ into
the symmetric and antisymmetric parts:
\be
\begin{cases}
f(z)=u(z)+v(z), \\
g(z)=u(z)-v(z).
\end{cases}
\ee
The system of the differential equations
(\ref{System_Bacher})
then takes the form
\be
\begin{cases}
u'(z)=2 u(z) v(z), \\
v'(z)=u^2(z)+v^2(z);
\label{Difeq_uv}
\end{cases}
\ \ \ u(0)=1; \ \ v(0)=0.
\ee
The first integral
(\ref{First_Int})
reads
\be
u^3(z)+3 u(z) v^2(z)=1.
\label{First_Int1}
\ee

Expressing
$v(z)$
from the first equation of
(\ref{Difeq_uv})
and substituting  it into the first integral
(\ref{First_Int1})
we get:
\be
\left[ u'(z) \right]^2= \frac{4}{3} u (1-u^3); \ \ \ u(0)=1.
\ee
From this it follows that
\be
u(z)= \frac{3 \wp(z; \, 0, \, \frac{4}{27})-1}{3 \wp(z; \, 0, \, \frac{4}{27})+2},
\ee
where $\wp(z; \, 0, \, \frac{4}{27})$ is the Weierstra{\ss} elliptic $\wp$-function
satisfying the master differential equation
\cite{Smirnov_VI}:
\be
\left[ \wp'(z; \, {\rm g}_2, \, {\rm g}_3 ) \right]^2=4 \wp^3(z; \, {\rm g}_2, \, {\rm g}_3 )-
{\rm g}_2 \wp(z; \, {\rm g}_2, \, {\rm g}_3 )
- {\rm g}_3
\label{WP_master_equation}
\ee
with the invariants ${\rm g}_2=0$, ${\rm g}_3=\frac{4}{27}$.

For $v(z)$
this results in
\be
v(z)= \frac{u'(z)}{2u(z)}=   \frac{9 \wp'(z; \, 0, \, \frac{4}{27})}{2 (3 \wp(z; \, 0, \, \frac{4}{27})+2)
(3 \wp(z; \, 0, \, \frac{4}{27})-1)}\,.
\ee
Finally,
\be
f(z)=u(z)+v(z)= \frac{18 \wp^2(z; \, 0, \, \frac{4}{27})+9 \wp'(z; \, 0, \, \frac{4}{27})
-12 \wp(z; \, 0, \, \frac{4}{27})+2 }{2 (3 \wp(z; \, 0, \, \frac{4}{27})+2) (3 \wp(z; \, 0, \, \frac{4}{27})-1)}.
\label{Bacher_WP}
\ee

The smallest positive real period of $f(z)$ turns
to be
\be
\omega_3 = 6r \equiv \pi_3 2^{-\frac{1}{3}},
\ee
where $\pi_3$ is defined in (\ref{pi3}).
Two complex periods can be chosen as
\be
\omega_{1,2}= e^{\pm i \frac{2\pi}{3}} \omega_3 =-3r \pm 3 \sqrt{3}r.
\ee

\begin{figure}[H]
 \begin{center}
  \epsfig{figure= 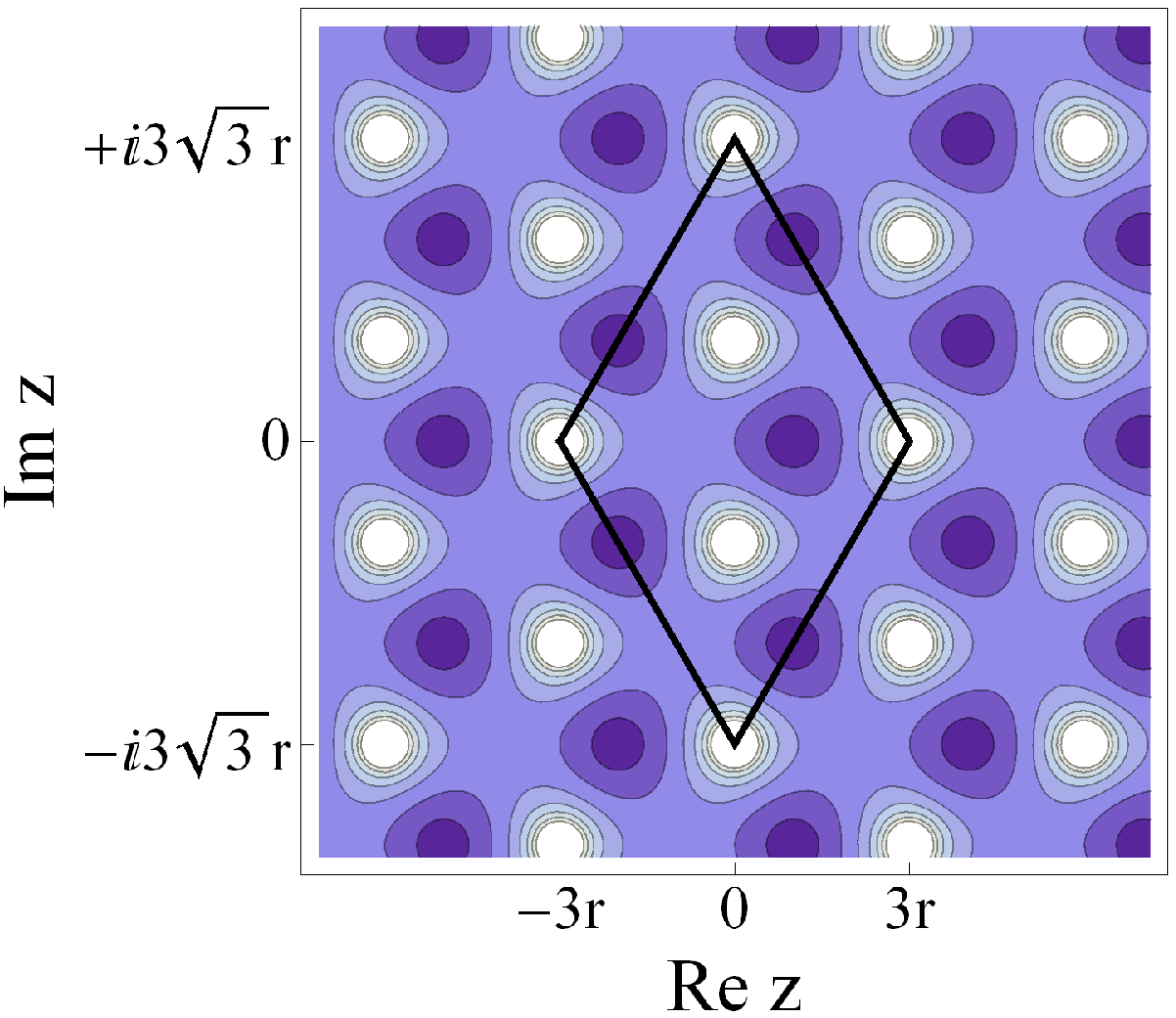 , height=8cm}
 \epsfig{figure= 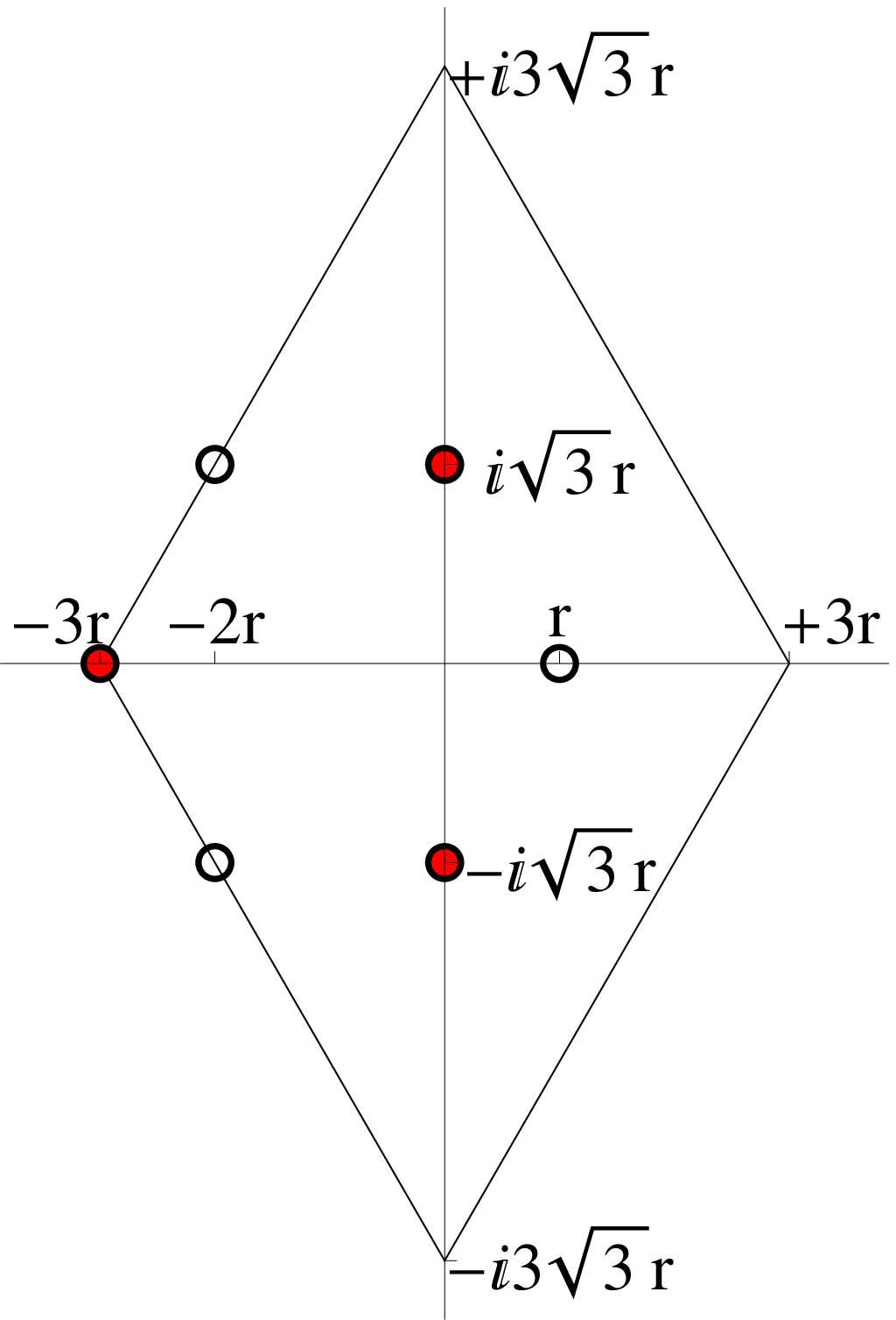 , height=8cm}
 \end{center}
  \caption{{\bf Left panel:} The level plot shows $|f(z)|$ (\ref{Pseud_Fact_f}) on the complex plane $z$.
  Black rhombus depicts the fundamental domain of $f(z)$ around the origin.
  {\bf Right panel:} Three poles (filled circles) and three zeroes (empty circles)
  of $f(z)$ in the complex plane $z$ inside the fundamental domain around the origin.}
\label{Fig_2}
  \end{figure}

Within the fundamental domain (see Fig.~\ref{Fig_2}) the function $f(z)$ possesses
$3$ simple poles:
\bi
\item a pole at $z=-3r$ with the residue $1$;
\item a pole at $z=i \sqrt{3} r$ with the residue $e^{-\frac{2 \pi i}{3}}$;
\item a pole at $z=-i \sqrt{3} r$ with the residue $e^{\frac{2 \pi i}{3}}$.
\ei
and $3$ zeros at $z= r$ and $z=-2r \pm i \sqrt{3} r$.

Quasi-renormalizable QFTs in which the result of LL-resummation  is
expressed through elliptic function look particularly appealing. The doubly periodic
structure of the Landau poles may reflect the highly non-trivial properties
of such theories. The physical examples of such theories are presented in Ref.~\cite{Bissex}.

\subsection{A More General Form of the Recurrence Relation}
\label{Sec_A0A1_equation}
\mbox

In this subsection we consider a particular case of the recurrence relation
(\ref{recrel_master1})
with arbitrary real parameters $A_0$, $A_1$ and $A_2=0$.
Our interest in considering this example relies on the
experience with the bissextile massless EFTs in $D=2$ \cite{Bissex}.
This recurrence relation turns to be equivalent to the following non-linear differential
functional equation for the generating function $f(z)$ (\ref{Gen_f_master}):
\be
f'(z)=A_0 f^2(z)+A_1 f^2(-z); \ \ f(0)=1.
\label{Case_from_Pade_difur}
\ee
We introduce the symmetric and antisymmetric parts of $f(z)$:
\be
u(z)= \frac{1}{2} \left( f(z)+f(-z) \right); \ \ \
v(z)= \frac{1}{2} \left( f(z)-f(-z) \right).
\ee
The equation
(\ref{Case_from_Pade_difur})
 then turns to be equivalent to the following system of differential equations:
\be
\begin{cases}
u'(z)=2(A_0-A_1) u(z) v(z), \\
v'(z)=(A_0+A_1)( u^2(z) + v^2(z));
\end{cases} \ \ \ u(0)=1; \ \ v(0)=0.
\label{Eq_uv_Smirnov}
\ee

First we consider the case
$A_0 \ne A_1$, $A_0 \ne -A_1$, $A_0 \ne 3A_1$.
One can check that the system of the differential equations
(\ref{Eq_uv_Smirnov})
possesses the following first integral
\be
\left(
u^2(z)+ \frac{3A_1-A_0}{A_0+A_1} v^2(z)
\right)^{A_0-A_1}=u^{A_0+A_1}(z).
\label{First_int_Smirnov}
\ee
Expressing
$v(z)$
from the first equation of
(\ref{Eq_uv_Smirnov})
and substituting it into the first integral
(\ref{First_int_Smirnov})
we obtain the following non-linear differential equation for
$u(z)$:
\be
\frac{3A_1-A_0}{4(A_0+A_1)(A_0-A_1)^2} \left[ u'(z) \right]^2= u^{N(A_0,A_1)}(z)-u^4(z); \ \ \ u(0)=1,
\label{Eq_u_elliptic_candidate}
\ee
where
\be
N(A_0,A_1)=\frac{3A_0-A_1}{A_0-A_1}.
\ee
In the case $N(A_0,A_1)=0,\,\ldots,\,3$ the solution of
(\ref{Eq_u_elliptic_candidate})
can be expressed through the elliptic functions (or their degeneracies). Below we
present the summary of the corresponding solutions.
\bi
\item In the case $N(A_0,A_1)=0 \Leftrightarrow A_1=3A_0$
the solution of  the equation
(\ref{Eq_uv_Smirnov})
can be expressed as follows
\be
&&
u(z)= \frac{\wp(z; \, -64 A_0^4,\,0)-4A_0^2}{\wp(z; \, -64 A_0^4,\,0)+4A_0^2}; \ \ \
v(z)= - \frac{u'(z)}{4A_0 u(z)}= - \frac{2A_0 \wp'(z; \, -64 A_0^4,\,0)}{\wp^2(z; \, -64 A_0^4,\,0)-16 A_0^4}
\nn \\ &&
\ee
through
the Weierstra{\ss} $\wp$-function
satisfying the master equation
(\ref{WP_master_equation})
with
${\rm g}_2=-64A_0^4$;
${\rm g}_3=0$.

\item The case $N(A_0,A_1)=1 \Leftrightarrow A_0=0$
corresponds to the generalization of (\ref{Diff_Eq_Bacher})
for arbitrary $A_1$. It is reduced to the former case for $A_1=-1$.

The solution can be expressed through  the Weierstra{\ss} $\wp$-function
\be
f(z)= \frac{18 \wp^2(z; \, 0,\, \frac{4A_1^6}{27} )
-9A_1 \wp'(z; \, 0,\, \frac{4A_1^6}{27} )-12 A_1^2 \wp(z; \, 0,\, \frac{4A_1^6}{27} )+2A_1^4 }
{2 (3 \wp(z; \, 0,\, \frac{4A_1^6}{27} )+2A_1^2) (3 \wp(z; \, 0,\, \frac{4A_1^6}{27} )-A_1^2)}.
\label{b_generalized_bacher}
\ee
The corresponding $\wp$-function satisfies
the master equation
(\ref{WP_master_equation})
with
${\rm g}_2=0$;
${\rm g}_3=\frac{4A_1^6}{27}$. Note that
(\ref{b_generalized_bacher})
indeed reduces to
(\ref{Bacher_WP})
for $A_1=-1$.

\item The case $N(A_0,A_1)=3 \Leftrightarrow A_1=0$ corresponds to the renormalizable theory
(\ref{Difeq_renormalizable})
with $b_1=A_0$.
\ei

We also present a short summary of solutions for the special cases  $A_0=A_1$,  $A_0=-A_1$, $A_0=3A_1$.
\bi
\item The case $A_0=A_1$ results in the non-trivial solution in terms of the
trigonometric functions:
\be
f(z)=1+\tan \left( 2 A_0 z \right).
\ee
\item The case $A_0=-A_1$ is trivial: $f(z)=1$.
\item In the case $A_0=3A_1$ the first integral of the system
(\ref{Eq_uv_Smirnov})
turns to be
\be
\log \left(  u(z) \right)= \frac{v^2(z)}{2 u^2(z)}.
\ee
The equation for $u(z)$ reads
\be
\left[ u'(z) \right]^2=32 A_1^2 u^4(z) \log \left( u(z) \right); \ \  u(0)=1.
\ee
The solution is therefore expressed in terms of the inverse of the Gauss error function
$${\rm erf}(x)=\frac{2}{\sqrt{\pi}} \int_0^x dy e^{-y^2}$$
\be
u(z)=e^{ \left( {\rm erf}^{-1}(4 \sqrt{\frac{2}{\pi }} A_1 z) \right)^2}; \ \ \
v(z)=  \frac{1}{4 A_1} \frac{d}{dz}   \left( {\rm erf}^{-1}(4 \sqrt{\frac{2}{\pi }} A_1 z) \right)^2.
\ee
\ei

In order to work out the additional solutions of (\ref{Eq_u_elliptic_candidate}) it is instructive
to perform the substitution
\be
w(z)=\frac{1}{u(z)}.
\ee
The equation (\ref{Eq_u_elliptic_candidate}) transforms into
\be
\frac{3A_1-A_0}{4(A_0+A_1)(A_0-A_1)^2} \left[ w'(z) \right]^2= w^{4-N(A_0,A_1)}(z)-1.
\ee
As is was already 
worked out, for
$N(A_0,A_1)=0,\,1,\, 2,\,3$
the solution is expressed in terms of elliptic functions (or their degeneracies).
For negative  integer $N(A_0,A_1)$ the solution will be expressed on terms of hyperelliptic functions
(or their degeneracies).

Another useful substitution is
\be
q(z)= \frac{1}{u^2(z)}.
\ee
The equation (\ref{Eq_u_elliptic_candidate}) then reads
\be
\frac{3A_1-A_0}{4(A_0+A_1)(A_0-A_1)^2} \left[ q'(z) \right]^2= q^{3-\frac{N(A_0,A_1)}{2}}(z)-q(z).
\ee
Apart from the already studied cases, this equation possesses the additional elliptic  solution for
$N(A_0,A_1)=-2 \Leftrightarrow A_0=\frac{3}{5}; \; A_1=1$:
\be
&&
u(z)=\sqrt{\frac{75 \wp \left(z; \,0,\,
4 \left( \frac{32}{75} \right)^3 \right) -64}{75 \wp \left(z; \,0,\, 4 \left( \frac{32}{75} \right)^3 \right) +32}}; \nn \\  &&
v(z)=- \frac{4500 \wp' \left(z; \,0,\,
4 \left( \frac{32}{75} \right)^3 \right)}{(75 \wp \left(z; \,0,\,
4 \left( \frac{32}{75} \right)^3 \right) -64)(75 \wp \left(z; \,0,\,
4 \left( \frac{32}{75} \right)^3 \right) +32)}\,,
\ee
where
$\wp \left(z; \,0,\,
4 \left( \frac{32}{75} \right)^3 \right)$
satisfies the master equation (\ref{WP_master_equation})
with
${\rm g}_2=0$;
${\rm g}_3=4 \left( \frac{32}{75} \right)^3$.


\section{Conclusions and Outlook}
\label{Sec_Concl}
\mbox

The summation of the leading 
logarithms in EFTs to all orders of loop expansion can provide us new valuable information
on the structure of the perturbation theory expansion in EFTs and help
to mirror the non-trivial properties of the general non-perturbative solution.

In this paper we address a class of non-linear recurrence equations that we
believe is relevant for the QFT applications. These recurrence relations
are equivalent to systems of non-linear differential equations for the
corresponding generating functions. We point out the explicit solutions
of these differential equations in terms of
elliptic functions (and their degeneracies).

We introduce the concept of quasi-renormalizable field theories in
which the generating function for the LL-coefficients is meromorphic
and may contain an infinite number of the Landau poles.
This can be seen as a generalization of the class of usual renormalizable
QFTs in which  the LL asymptotic behavior is controlled by the presence of
a single Landau pole. The situation in which the LL asymptotic behavior of EFT
is described by a (doubly)periodic function is particularly interesting.
The periodic pole structure 
could have relation
to the mass spectrum of the full theory.
The explicit examples of quasi-renormalizable effective field theories
in $D=2$ with an infinite number of periodically located Landau poles
are presented in Ref.~\cite{Bissex}.

We also expect possible non-trivial connections between QFT and
various branches of pure mathematics such as combinatorics,
theory of  continued fractions and theory of orthogonal polynomials.

\section*{Acknowledgements}
\mbox

We are grateful to   D. Kazakov, N. Kivel, J. Linzen, and N. Sokolova
for many enlightening and inspiring discussions
and to R. Bacher for correspondence.
We also owe sincere thanks to V.B.~Matveev for help and advise.
KMS is also grateful to the warm hospitality of the TP II group of
Ruhr-Universit\"at Bochum, where the essential part of this study was done.

The  work of MVP and KMS was supported by the grant CRC110 (DFG).
AOS acknowledges the support by the RFBR grant 18-51-18007.


\begin{thebibliography}{99}

\bibitem{Vasiliev_green_Book}
A.~N.~Vasiliev,
{\it ``The Field Theoretic Renormalization Group in Critical Behavior Theory and Stochastic Dynamics''}
(Chapman \& Hall/ CRC, 2004).

\bibitem{Colangelo:1995np}
  G.~Colangelo,
  \href{https://www.sciencedirect.com/science/article/pii/037026939501162J?via\%3Dihub}{Phys.\ Lett.\ B {\bf 350}, 85 (1995)},
  Erratum: [Phys.\ Lett.\ B {\bf 361}, 234 (1995)]
  [\href{https://arxiv.org/abs/hep-ph/9502285}{hep-ph/9502285}].

\bibitem{Kazakov:1987jp}
  D.~I.~Kazakov,
\href{https://link.springer.com/article/10.1007\%2FBF01017179}{Theor.\ Math.\ Phys.\  {\bf 75}, 440 (1988)}
  [Teor.\ Mat.\ Fiz.\  {\bf 75}, 157 (1988)].
 %

\bibitem{Buchler:2003vw}
  M.~Buchler and G.~Colangelo,
  %
  \href{https://link.springer.com/article/10.1140\%2Fepjc\%2Fs2003-01390-2}{Eur.\ Phys.\ J.\ C {\bf 32}, 427 (2003)}
  [\href{http://arxiv.org/abs/hep-ph/0309049}{hep-ph/0309049}].

\bibitem{Bissegger:2006ix}
  M.~Bissegger, and A.~Fuhrer,
   \href{https://doi.org/10.1016/j.physletb.2007.01.025}{Phys.\ Lett.\ B {\bf 646}, 72 (2007)}
  [\href{http://arxiv.org/abs/hep-ph/0612096}{hep-ph/0612096}].

\bibitem{Kivel:2008mf}
  N.~Kivel, M.~V.~Polyakov, and A.~Vladimirov,
  %
  \href{https://doi.org/10.1103/PhysRevLett.101.262001}{Phys.\ Rev.\ Lett.\  {\bf 101}, 262001 (2008)}
  [\href{http://arxiv.org/abs/arXiv:0809.3236}{arXiv:0809.3236 [hep-ph]}].


\bibitem{Koschinski:2010mr}
  J.~Koschinski, M.~V.~Polyakov, and A.~A.~Vladimirov,
  %
  \href{https://doi.org/10.1103/PhysRevD.82.014014}{Phys.\ Rev.\ D {\bf 82}, 014014 (2010)}
  [\href{http://arxiv.org/abs/arXiv:1004.2197}{arXiv:1004.2197 [hep-ph]}].

\bibitem{Kivel:2009az}
  N.~A.~Kivel, M.~V.~Polyakov, and A.~A.~Vladimirov,
  %
  \href{https://link.springer.com/article/10.1134/S0021364009110022}{JETP Lett.\  {\bf 89}, 529 (2009)}
  [\href{http://arxiv.org/abs/arXiv:0904.3008}{arXiv:0904.3008 [hep-ph]}].

\bibitem{VladimirovThesis}
A.~A.~Vladimirov,
{\it Infrared Logarithms in Effective Field Theories}, PhD thesis, Ruhr University, Bochum, 2010, unpublished \\
\href{http://www-brs.ub.ruhr-uni-bochum.de/netahtml/HSS/Diss/VladimirovAlexey/diss.pdf}{{\small http://www-brs.ub.ruhr-uni-bochum.de/netahtml/HSS/Diss/VladimirovAlexey/diss.pdf}}.

\bibitem{Polyakov:2010pt}
  M.~V.~Polyakov, and A.~A.~Vladimirov,
  %
 \href{https://link.springer.com/article/10.1007/s11232-011-0126-7}{Theor.\ Math.\ Phys.\  {\bf 169}, 1499 (2011)}
  [\href{http://arxiv.org/abs/arXiv:1012.4205}{arXiv:1012.4205 [hep-th]}].


\bibitem{Ananthanarayan:2018kly}
  B.~Ananthanarayan, S.~Ghosh, A.~Vladimirov, and D.~Wyler,
 %
  \href{https://link.springer.com/article/10.1140\%2Fepja\%2Fi2018-12555-9}{ Eur.\ Phys.\ J.\ A {\bf 54}, no. 7, 123 (2018)}
  [\href{http://arxiv.org/abs/arXiv:1803.07013}{arXiv:1803.07013 [hep-ph]}].


\bibitem{Perevalova:2011qi}
  I.~A.~Perevalova, M.~V.~Polyakov, A.~N.~Vall, and A.~A.~Vladimirov,
  {\it ``Chiral Inflation of the Pion Radius,''}
 \href{http://arxiv.org/abs/arXiv:1105.4990}{arXiv:1105.4990 [hep-ph]}.




\bibitem{Bijnens:2009zi}
  J.~Bijnens, and L.~Carloni,
 %
\href{https://www.sciencedirect.com/science/article/pii/S0550321309005732?via\%3Dihub}{ Nucl.\ Phys.\ B {\bf 827}, 237 (2010)}
  [\href{http://arxiv.org/abs/arXiv:0909.5086}{arXiv:0909.5086 [hep-ph]}].


\bibitem{Bijnens:2010xg}
  J.~Bijnens, and L.~Carloni,
 %
  \href{https://www.sciencedirect.com/science/article/pii/S0550321310004980?via\%3Dihub}{ Nucl.\ Phys.\ B {\bf 843}, 55 (2011)}
  [\href{http://arxiv.org/abs/arXiv:1008.3499}{arXiv:1008.3499 [hep-ph]}].

\bibitem{Bijnens:2014ila}
  J.~Bijnens, and A.~A.~Vladimirov,
  \href{https://www.sciencedirect.com/science/article/pii/S0550321314003939?via\%3Dihub}{Nucl.\ Phys.\ B {\bf 891}, 700 (2015)}
  [\href{http://arxiv.org/abs/arXiv:1409.6127}{arXiv:1409.6127 [hep-ph]}].

\bibitem{Bissex}
J.~Linzen, M.~V.~Polyakov, K.~M.~Semenov-Tian-Shansky, and
N.~S.~Sokolova, {\it Exact Summation of Leading Infrared Logarithms in $2D$ Effective Field Theories}, {under preparation}.


\bibitem{Landau}
L. D. Landau,
A. A. Abrikosov,
and I. M. Khalatnikov,
Dokl. Akad. Nauk SSSR {\bf 95},   497, 773, 1177  (1954).




\bibitem{Malyshev:2003py}
  D.~Malyshev,
  \href{https://www.sciencedirect.com/science/article/pii/S0370269303015958?via\%3Dihub}{Phys.\ Lett.\ B {\bf 578}, 231 (2004)}
 [\href{http://arxiv.org/abs/hep-th/0307301}{hep-th/0307301}].


\bibitem{Malyshev:2004ax}
  D.~V.~Malyshev,
 \href{https://link.springer.com/article/10.1007\%2Fs11232-005-0086-x}{Theor.\ Math.\ Phys.\  {\bf 143}, 505 (2005)}
  [Teor.\ Mat.\ Fiz.\  {\bf 143}, 22 (2005)]
  [\href{http://arxiv.org/abs/hep-th/0408230}{hep-th/0408230}].

\bibitem{Kreimer:1997dp}
  D.~Kreimer,
  \href{https://www.intlpress.com/site/pub/pages/journals/items/atmp/content/vols/0002/0002/a004/index.html}{Adv.\ Theor.\ Math.\ Phys.\  {\bf 2}, 303 (1998)}
  [\href{http://arxiv.org/abs/q-alg/9707029}{q-alg/9707029}].



\bibitem{Borlakov:2016mwp}
  A.~T.~Borlakov, D.~I.~Kazakov, D.~M.~Tolkachev, and D.~E.~Vlasenko,
  \href{https://link.springer.com/article/10.1007\%2FJHEP12\%282016\%29154}{JHEP {\bf 1612}, 154 (2016)}
  [\href{http://arxiv.org/abs/arXiv:1610.05549}{arXiv:1610.05549 [hep-th]}].

\bibitem{Kazakov:2016wrp}
  D.~I.~Kazakov, and D.~E.~Vlasenko,
%
  \href{https://doi.org/10.1103/PhysRevD.95.045006}{Phys.\ Rev.\ D {\bf 95}, no. 4, 045006 (2017)}
  [\href{http://arxiv.org/abs/arXiv:1603.05501}{arXiv:1603.05501 [hep-th]}].


\bibitem{Borlakov:2017vra}
  D.~I.~Kazakov, A.~T.~Borlakov, D.~M.~Tolkachev, and D.~E.~Vlasenko,
  \href{https://doi.org/10.1103/PhysRevD.97.125008}{Phys.\ Rev.\ D {\bf 97}, no. 12, 125008 (2018)}
  [\href{http://arxiv.org/abs/arXiv:1712.04348}{arXiv:1712.04348 [hep-th]}].

\bibitem{Julia}
J.~ Koschinski,
{\it Leading Logarithms
in Four Fermion Theories},
PhD thesis, Ruhr University, Bochum, 2015, unpublished \\
\href{https://d-nb.info/1109051174/34}{{\small https://d-nb.info/1109051174/34}}.


\bibitem{Hammerstein}
A. Hammerstein, 
Acta Math. {\bf 54},  117 (1930).

\bibitem{Dolph}
C.~L.~Dolph,
Trans. Amer. Math. Soc., Vol. {\bf 66}, 
289 (1949).


\bibitem{Oesis_Cat}
N.~J.~A.~Sloane, {\it The On-Line Encyclopedia of Integer Sequences (OEIS)}
\href{https://oeis.org/A000108}{{\small https://oeis.org/A000108}}.


\bibitem{Migdal:1977nu}
  A.~A.~Migdal,
  \href{https://www.sciencedirect.com/science/article/pii/0003491677901816}{Annals Phys.\  {\bf 109}, 365 (1977)}.



\bibitem{Migdal:1977ut}
  A.~A.~Migdal,
  \href{https://www.sciencedirect.com/science/article/pii/0003491678901410?via\%3Dihub}{Annals Phys.\  {\bf 110}, 46 (1978)}.

\bibitem{Migdal:2011pi}
  A.~A.~Migdal,
  {\it Meromorphization of Large $N$ QFT},
  \href{http://arxiv.org/abs/arXiv:1109.1623}{arXiv:1109.1623 [hep-th]}.

\bibitem{Migdal:2011gf}
  A.~A.~Migdal,
 {\it Integral Equation for CFT/String Duality},
  \href{http://arxiv.org/abs/arXiv:1107.2370}{arXiv:1107.2370 [hep-th]}.

\bibitem{Erlich:2006hq}
  J.~Erlich, G.~D.~Kribs and I.~Low,
  %
  \href{https://journals.aps.org/prd/abstract/10.1103/PhysRevD.73.096001}{Phys.\ Rev.\ D {\bf 73}, 096001 (2006)}
  [\href{https://arxiv.org/abs/hep-th/0602110}{hep-th/0602110}].



\bibitem{Baher}
R.~Bacher, and P.~Flajolet,
%
%
\href{https://doi.org/10.1007/s11139-009-9186-9}{Ramanujan J {\bf 21}, 71 (2010)}
[\href{https://arxiv.org/abs/0901.1379}{arXiv:0901.1379 [math]}].

\bibitem{Flajolet}
E.
~van Fossen Conrad, and P.
~Flajolet, 
\href{https://www.mat.univie.ac.at/~slc/wpapers/s54conflaj.pdf}{S\'eminaire Lotharingien de Combinatoire, B54g, 1 (2006)} [\href{https://arxiv.org/abs/math/0507268}{arXiv:math/0507268 [math.CO]}].


\bibitem{Oesis_Pf}
N.~J.~A.~Sloane, {\it The On-Line Encyclopedia of Integer Sequences (OEIS)}
\href{https://oeis.org/A098777}{{\small https://oeis.org/A098777}}.



\bibitem{Dixon}
A. C. Dixon, 
Quart. J. Math., 
{\bf 24},  p. 167-233 (1890).

\bibitem{Smirnov_VI} В. И. Смирнов, <<Курс Высшей Математики>>, Том 3, Часть 2, <<Наука>>, Москва, 1974
(in Russian).



\end{thebibliography}
\end{document}